\documentclass[12pt]{iopart}
\usepackage{graphicx}
\begin{document}

\title[Superconductivity up to 30 K in BaFe$_{2}$(As$_{1-x}$P$_{x}$)$_{2}$]{Superconductivity up to 30 K in the vicinity of quantum critical point in BaFe$_{2}$(As$_{1-x}$P$_{x}$)$_{2}$}

\author{Shuai Jiang$^1$, Hui Xing$^1$, Guofang Xuan$^1$, Cao Wang$^1$, Zhi Ren$^1$, Chunmu Feng$^2$, Jianhui Dai$^1$, Zhu'an Xu$^1$, Guanghan Cao$^1$\footnote[3]{Corresponding author. E-mail
address: ghcao@zju.edu.cn}}

\address{$^1$ Department of Physics, Zhejiang University, Hangzhou 310027, China}

\address{$^2$ Test {\&} Analysis Center, Zhejiang University,
Hangzhou 310027, China}

\begin{abstract}
We report bulk superconductivity induced by an isovalent doping of
phosphorus in BaFe$_{2}$(As$_{1-x}$P$_{x}$)$_{2}$. The P-for-As
substitution results in shrinkage of lattice, especially for the
FeAs block layers. The resistivity anomaly associated with the
spin-density-wave (SDW) transition in the undoped compound is
gradually suppressed by the P doping. Superconductivity with the
maximum $T_c$ of 30 K emerges at $x$=0.32, coinciding with a
magnetic quantum critical point (QCP) which is evidenced by the
disappearance of SDW order and the linear temperature-dependent
resistivity in the normal state. The $T_c$ values were found to
decrease with further P doping, and no superconductivity was
observed down to 2 K for $x\geq$ 0.77. The appearance of
superconductivity in the vicinity of QCP hints to the
superconductivity mechanism in iron-based arsenides.

\ (Some figures in this article are in colour only in the electronic
version)

\end{abstract}


\submitto{\JPCM}as a FAST TRACK COMMUNICATIONS


\maketitle

\section{Introduction}
The discovery of high-temperature superconductivity in iron
arsenides\cite{Kamihara08,chen-nature} was an exciting breakthrough
in the year 2008,\cite{science} which has led to a number of new
superconductors\cite{122,111,11} all containing antifluorite-like
FeAs or FeSe layers. In most cases, superconductivity was induced by
doping charge carriers into a stoichiometric parent compound which
shows a collinear antiferromagnetic spin-density-wave ground
state\cite{daiPC}. The chemical doping was initially carried out
only outside the FeAs layers,\cite{Kamihara08,122,Th,Sr} with the
electronic phase diagram\cite{CXH,Zhao,Gooch} resembling that of
cuprate superconductors. Later, however, it was found that the even
doping at the Fe site (with Co\cite{Co} or Ni\cite{Ni}) could induce
superconductivity. Besides, superconductivity was able to be
stabilized in the parent compounds simply by applying hydrostatic
pressures.\cite{P-Ca122} The latter two results contrast sharply
with those in cuprates.

Very recently, we observed superconductivity at 26 K in
EuFe$_2$As$_2$ by the partial substitution of arsenic with
phosphorus.\cite{ren} This isovalent substitution does not introduce
additional electrons or holes into the system, thus it can be
roughly understood in terms of chemical pressures. Owing to the
ferromagnetic ordering of Eu$^{2+}$ moments below 20 K coexisted
with superconductivity, the Meissner effect was not obvious.
Subsequently, we found that the P/As substitution induced
superconductivity in a prototype "1111" material LaFeAsO, but the
superconducting critical temperature $T_c$ is only $\sim$10
K,\cite{Wang2009} which is substantially lower than the $T_c$ value
under physical pressures\cite{Okada}. Here we employ the same doping
strategy on another prototype "122" parent compound
BaFe$_{2}$As$_{2}$\cite{Rotter}. Bulk superconductivity up to 30 K
with strong Meissner diamagnetism has been observed for $x$=0.32.
Our systematic measurements of resistivity and magnetization reveal
an electronic phase diagram characterized by a dome-like
superconducting region located near a magnetic quantum critical
point (QCP). The present work is consistent with a recent
theoretical prediction which shows a unique type of QCP resulting
from a competition between electronic localization and
itinerancy.\cite{DaiJH}

\section{Experimental}

Polycrystalline samples of BaFe$_{2}$(As$_{1-x}$P$_{x}$)$_{2}$ (with
the nominal P content $x$=0, 0.1, 0.2, 0.25, 0.3, 0.35, 0.4, 0.45,
0.5, 0.6, 0.7, 0.8 and 0.9 and 1) were synthesized by solid state
reaction with the high-purity elements Ba, Fe, As and P. Mixture of
the elements in the stoichiometric ratio was pressed into pellets in
an argon-filled glove-box. The pellets were sealed in evacuated
quartz tubes and annealed at 600 K for 12 h, then at 1173 K for 40
h. The resultant was reground, pelletized, and annealed in evacuated
quartz tubes at 1273 K for 24 h. Finally the samples were cooled
down to room temperature by shutting off the furnace.

Powder x-ray diffraction (XRD) was performed at room temperature
using a D/Max-rA diffractometer with Cu-K$_{\alpha}$ radiation and a
graphite monochromator. The structural refinement was performed
using the programme RIETAN 2000.\cite{Izumi} The electrical
resistivity was measured using a standard four-probe method. The
measurements of dc magnetization were performed down to 1.8 K on a
Quantum Design Magnetic Property Measurement System (MPMS-5). Both
the zero-field-cooling (ZFC) and field-cooling (FC) protocols were
employed under the magnetic field of 10 Oe.

\section{Results and discussion}

BaFe$_{2}$(As$_{1-x}$P$_{x}$)$_{2}$ system crystallizes in
ThCr$_2$Si$_2$-type structure with the space group I4/\emph{mmm}.
Except the main phase, minor secondary phase was identified as
Fe$_{2}$P. Figure 1(a) shows the XRD pattern and its Rietveld
refinement profile of a typical BaFe$_{2}$(As$_{1-x}$P$_{x}$)$_{2}$
sample with $x$=0.35 at room temperature. The two-phase Rietveld
refinement is successful, as indicated by the fairly good
reliability factors $R_{wp}$=8.5\% and $S$=1.8. It was found that
the fitted P content is a little lower than the nominal one,
consistent with the existence of impurity Fe$_{2}$P (about 10\% in
molar ratio). The refined lattice parameters plotted as functions of
$x$ are shown in Fig. 1(b). The cell volume decreases monotonically
with the P doping. We notice that, for the samples with no
significant Fe$_{2}$P impurity (\emph{e.g.}, $x$=0.45 and 0.9), the
data points of the cell volume fall exactly on the straight line, in
accordance with the Vegard's law. Therefore, the actual P content
can be estimated from the measured cell volume, which gives the
corrected P content $x'$=0.02, 0.13, 0.2, 0.24, 0.32, 0.36, 0.47,
0.56, 0.63 and 0.77 for the nominal $x$ values of 0.1, 0.2, 0.25,
0.3, 0.35, 0.4, 0.5, 0.6, 0.7 and 0.8, respectively. This correction
is basically (within the uncertainty of $\pm$0.02) consistent with
the Rietveld refinement results.

\begin{figure}
\center
\includegraphics[width=8cm]{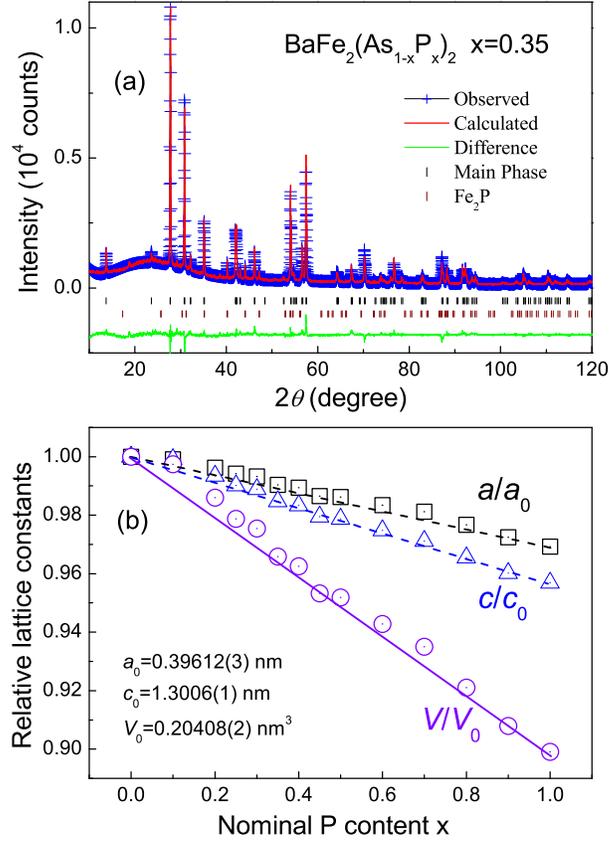}
\caption{(a) X-ray powder diffraction pattern and its Rietveld
refinement profile for the BaFe$_{2}$(As$_{1-x}$P$_{x}$)$_{2}$
($x$=0.35) sample. The secondary minor phase is identified as
Fe$_{2}$P. (b) Relative lattice constants (compared with the undoped
compound) as functions of nominal P content $x$. The solid straight
line obeys the Vegard's law.}
\end{figure}

Table 1 compares the crystal structures between the undoped and
P-doped ($x'$=0.32) BaFe$_{2}$As$_{2}$. A remarkable difference is
seen from the position of As/P. The P doping makes the pnictogen
atoms move toward the iron sheets, which leads to a dramatic
decrease of FeAs-layer thickness. This explains why the decrease in
$c$-axis is more pronounced (compared with the $a$-axis), as shown
in Fig. 1(b). In fact, the spacing of FeAs layers decreases little
with P doping. This indicates that P doping impacts mainly on the
FeAs block layers. As a result, the bond angle of As-Fe-As along
[110] directions increases substantially.

\begin{table}
\caption{\label{tab:table1}Comparison of room-temperature crystal
structures between the undoped and P-doped ($x'$=0.32)
BaFe$_{2}$As$_{2}$ with the space group I$4/mmm$. The atomic
coordinates are as follows: Ba (0, 0, 0); Fe (0.5, 0, 0.25); As/P
(0, 0, $z$).}

\center
\begin{tabular}{lcr}
\hline
Compounds&BaFe$_{2}$As$_{2}$&BaFe$_{2}$(As$_{0.68}$P$_{0.32}$)$_{2}$\\
\hline
$a$ ({\AA}) & 3.9612(3) &3.9231(3)\\
$c$ ({\AA}) & 13.006(1) &12.805(1)\\
$V$ ({\AA}$^3$) & 204.08(2) & 197.08(2)\\
$z$ of As/P & 0.3546(2) & 0.3530(2)\\
FeAs-layer thickness ({\AA}) & 2.721(2) &2.638(2)\\
FeAs-layer spacing ({\AA}) & 3.782(2) &3.765(2)\\
As-Fe-As angle ($^{\circ}$) & 111.0(1) &112.2(1)\\
\hline
\end{tabular}

\end{table}

Figure 2 shows the temperature dependence of resistivity for
BaFe$_{2}$(As$_{1-x}$P$_{x}$)$_{2}$. For the undoped
BaFe$_{2}$As$_{2}$, $\rho$ drops rapidly below 140 K, consistent
with previous report\cite{Rotter}. The resistivity anomaly was
revealed to be associated with a simultaneous structural and
spin-density-wave (SDW) transition.\cite{Bao} Upon doping with P,
the transition is suppressed, exhibiting the decrease of transition
temperature ($T_{SDW}$) and the weakening of the resistivity drop.
For example, $T_{SDW}$ is suppressed to $\sim$ 90 K for the sample
of $x'$=0.13, where only a resistivity kink appears. When the P
content increases to 20\%, a sudden decrease in resistivity is seen
below 10 K in addition to an observable kink at $\sim$ 70 K. The
resistivity drop at lower temperature is due to a superconducting
transition (confirmed by the magnetic measurements below). In the
intermediate doping range (0.24$\leq x\leq$0.63), superconductivity
with higher $T_c$s was observed. The maximum transition temperature
is 30 K for $x'$=0.32. The resistive transition tail is probably due
to the influence of grain boundaries and/or minor ferromagnetic
Fe$_2$P impurity, rather than sample inhomogeneity, since nearly
perfect diamagnetism was measured (see below). For the higher P
doping ($x\geq$0.77), no superconductivity was observed down to 3 K.
One notes that the residual resistance ratio (RRR), defined by the
ratio of the resistance at room temperature to the residual
resistance at very low temperature, is 20, 35 and 75 for $x'$=0.77,
0.9 and 1.0, respectively. The high RRR values suggest high quality
of the samples.

\begin{figure}
\center
\includegraphics[width=8cm]{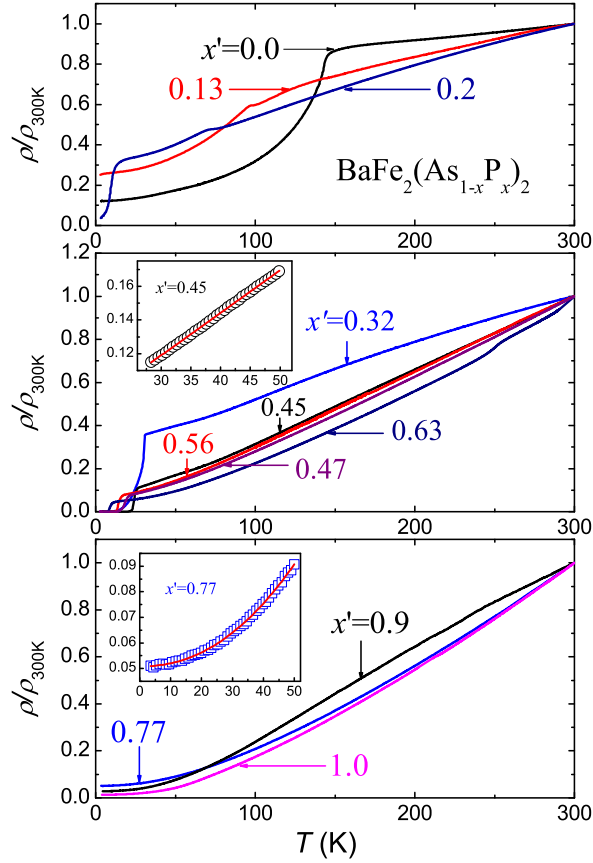}
\caption{Temperature dependence of resistivity for
BaFe$_{2}$(As$_{1-x'}$P$_{x'}$)$_{2}$ samples. The data are
normalized for the convenience of comparison. All the curves are
labeled with the corrected P content. The insets display expanded
plot for showing the fit of the normal-state resistivity with the
formula $\rho(T)=\rho_{0}+\alpha T^{n}$. }
\end{figure}

To confirm superconductivity, the temperature dependence of dc
magnetic susceptibility for BaFe$_{2}$(As$_{1-x}$P$_{x}$)$_{2}$ was
measured, as shown in Fig. 3. The sample of $x'$=0.2 shows strong
diamagnetic transition below 7 K, consistent with the drop in
resistivity below 10 K shown in Fig. 2. In spite of the large
superconducting fraction, however, it is difficult to ascertain
whether SDW order coexists microscopically with superconductivity or
not. For $x$=0.32, a stronger diamagnetic signal is observed below
30 K. The magnetic shielding is even beyond the ideal value of 100\%
apparently, since the demagnetization effect is not included here.
The Meissner volume fraction achieves $\sim$ 30\%, demonstrating the
bulk nature of superconductivity. Other superconducting samples also
displays similar bulk property.

\begin{figure}
\center
\includegraphics[width=8cm]{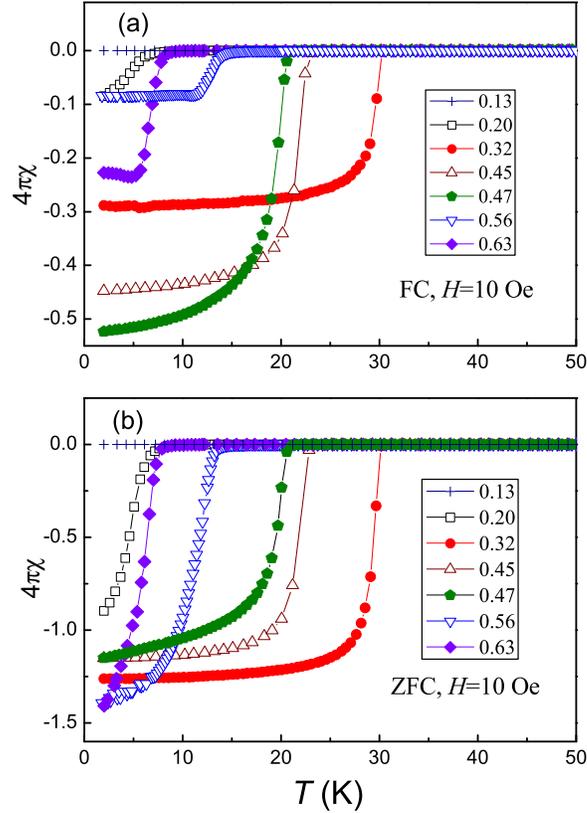}
\caption{Superconducting diamagnetic transitions in
BaFe$_{2}$(As$_{1-x'}$P$_{x'}$)$_{2}$. Note that the magnetic
background coming from the ferromagnetic impurity of Fe$_2$P has
been removed. No correction of demagnetization has been made.}
\end{figure}

As is known, the normal-state resistivity contains valuable
information on electronic scattering and even superconducting
mechanisms. Dense polycrystalline samples may show intrinsic
transport property because the grain boundaries scatter electrons
like impurities. Thus the following analysis is valid even though
our data were obtained from polycrystalline samples. Assuming that
the intrinsic scattering produces a power-law resistivity at low
temperatures, the total resistivity can be expressed as
$\rho(T)=\rho_{0}+\alpha T^{n}$, where $\rho_0$ represents extrinsic
resistivity coming from the impurity/defect scattering. By fitting
the $\rho(T)$ data in the range of $T_c< T\leq$50 K, the exponent
$n$ was obtained. The inset of Fig. 4 shows that most samples have
the $n$-value close to 2.0, suggesting Fermi-liquid state due to
electron-electron interactions. Interestingly, $n$ is close to unity
(i.e., linear $T$-dependent normal-state resistivity) for the
samples with $x'$=0.3$\sim$0.45. The linear low-temperature
resistivity belongs to a non-Fermi liquid behaviour, as usually
observed in the regime of quantum criticality. Remember that the SDW
order is suppressed with the P doping, we argue that there exists a
QCP at $x'\sim$1/3 in the BaFe$_{2}$(As$_{1-x}$P$_{x}$)$_{2}$
system. This result is quite consistent with a recent theoretical
prediction which shows a unique type of QCP resulting from a
competition between electronic localization and
itinerancy\cite{DaiJH}. We aware that similar QCP was also proposed
in (K,Sr)Fe$_{2}$As$_{2}$\cite{Gooch} system.

The electronic phase diagram is concluded in Fig. 4. The undoped
BaFe$_2$As$_2$ has the ground state of SDW order. With the P doping,
$T_{SDW}$ decreases gradually, forming a phase region of SDW. The
suppression of SDW can be easily explained in terms of $J_{1}-J_{2}$
model\cite{Si,Yildirim}, because $J_{2}$ is expected to decrease
upon P doping. On the other hand, the scenario of Fermi surface
nesting\cite{Dong} is not quite straightforward to meet the
experimental observation. In the vicinity of above-mentioned QCP, an
asymmetric superconducting (SC) dome emerges. The optimal doping
level (with the maximum $T_c$) coincides with the QCP, where a
non-Fermi-liquid behviour shows up. This strongly suggests that spin
fluctuations play an important role for the superconductivity. In
the overlapped area of SDW and SC, it is not clear whether SDW
coexists microscopically with SC at this moment. The upper-right
metallic region displays Fermi-liquid-like state, except that
non-Fermi-liquid behaviour appears near the QCP. This crossover in
electronic excitation states resembles those of
(K,Sr)Fe$_{2}$As$_{2}$\cite{Gooch} and
Ba(Fe$_{1-x}$Co$_{x}$)$_{2}$As$_{2}$\cite{PD122Co} systems.

\begin{figure}
\center
\includegraphics[width=8cm]{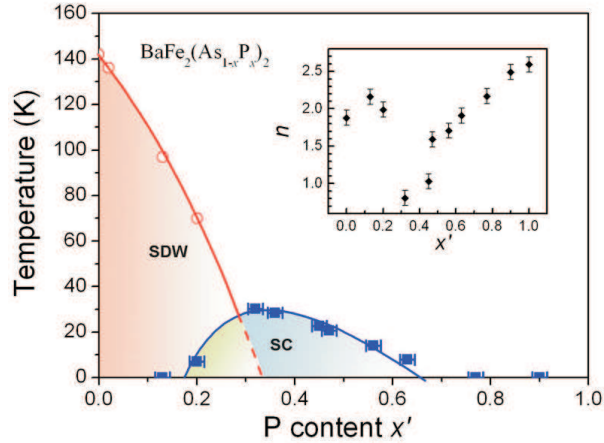}
\caption{Electronic phase diagram in
BaFe$_{2}$(As$_{1-x'}$P$_{x'}$)$_{2}$. SDW and SC denote
spin-density-wave and superconductivity, respectively. The inset
shows the variations of the exponent $n$ in the formula
$\rho(T)=\rho_{0}+\alpha T^{n}$, obtained by the fitting of
normal-state resistivity at low temperatures.}
\end{figure}

It was reported that BaFe$_2$As$_2$ became superconducting up to 29
K under the high pressure of $\sim$4 GPa.\cite{P-SrBa122} The $T_c$
value is very close to the present one in
BaFe$_{2}$(As$_{1-x}$P$_{x}$)$_{2}$. This implies that chemical
pressure generated by P doping acts in a similar way to the physical
pressure. Meanwhile, the disorder induced by the P doping hardly
influences superconductivity. The relatively low $T_c^{max}$ value
(compared with 38 K in Ba$_{0.6}$K$_{0.4}$Fe$_2$As$_2$\cite{122})
can be well understood by an empirical structural rule for $T_c$
variations.\cite{Zhao} A very recent work\cite{Kimber} also reveals
similarities between structural distortions under pressure and
chemical doping in BaFe$_2$As$_2$. Therefore, it is the structural
details of FeAs layers (corresponding to specific electronic
structures), rather than charge doping level, that controls the
ground states in iron arsenides.

To summarize, BaFe$_2$As$_2$ has been tuned superconducting through
an isovalent doping of phosphorus at the arsenic site, for the first
time. Unlike previous doping strategy with changing the
concentration of charge carrier, the present P doping does not alter
the number of valence electrons. Structural measurements indicates
that the chemical pressure generated by the P doping impacts mainly
on the FeAs layers. The P-for-As substitution suppresses the SDW
ordering, and favors superconductivity, similarly to phenomena of
the carrier doping\cite{122,CXH,Zhao}. A magnetic QCP is evidenced
simultaneously by the disappearance of SDW ordering and the
non-Fermi-liquid behaviour. The maximum $T_c$ of 30 K is observed
where the QCP is located. Theoretical studies\cite{Moriya} indicate
that superconductivity around magnetic QCP is associated with the
spin fluctuation mechanism. Thus the present result suggests that
spin fluctuations play crucial role for the superconductivity in
iron-based arsenides.

\section*{Acknowledgments}
This work is supported by the NSF of China (No. 10674119), National
Basic Research Program of China (No. 2007CB925001) and the PCSIRT of
the Ministry of Education of China (IRT0754).


\section*{References}
\begin{thebibliography}{10}
\bibitem{Kamihara08}Kamihara Y, Watanabe T, Hirano M and Hosono H 2008
J. Am. Chem. Soc. \textbf{130} 3296
\bibitem{chen-nature}Chen X H, Wu T, Wu G, Liu R H, Chen H and Fang D F 2008
Nature \textbf{453} 761
\bibitem{science}The Editorial 2008 \emph{Breakthrough of the year} Science \textbf{322} 1770
\bibitem{122}Rotter M, Tegel M and Johrendt D 2008 Phys. Rev. Lett. \textbf{101} 107006
\bibitem{111}Wang W C \emph{et al} 2008 Solid Sate Communications \textbf{148}
538; Pitcher M J \emph{et al} 2008 Chem. Commun. 5918; Tapp J H
\emph{et al} 2008 Phys. Rev. B \textbf{78} 060505(R)
\bibitem{11}Hsu F C \emph{et al} 2008 Proc. Natl. Acad. Sci. \textbf{105} 14262
\bibitem{daiPC}Cruz C \emph{et al} 2008 Nature \textbf{453} 899
\bibitem{Th}Wang C \emph{et al} 2008 Europhysics Lett. \textbf{83} 67006
\bibitem{Sr}Wen H H \emph{et al} 2008 Europhysics Lett. \textbf{82} 17009
\bibitem{CXH}Liu R H \emph{et al} 2008 Phys. Rev. Lett. \textbf{101} 087001
\bibitem{Zhao}Zhao J \emph{et al} 2008 Nat. Mater. \textbf{7} 953
\bibitem{Gooch}Gooch M, Bing L, Lorenz B, Guloy A M and Chu C W 2009
Phys. Rev. B \textbf{79} 104504
\bibitem{Co}Sefat A S \emph{et al.} 2008 \PR B \textbf{78} 104505; Wang C \emph{et al} 2009 \PR B \textbf{79} 054521
\bibitem{Ni}Cao G H \emph{et al} 2009 \PR B \textbf{79} 174505; Li L J \emph{et al} 2009 New J. Phys. \textbf{11} 025008
\bibitem{P-Ca122}Torikachvili M S \emph{et al} 2008 Phys. Rev. Lett. \textbf{101} 057006; Park T \emph{et al} 2008 J. Phys.: Condensed Matter \textbf{20} 322204
\bibitem{ren}Zhi R \emph{et al} 2009 Phys. Rev. Lett. \textbf{102} 137002
\bibitem{Wang2009}Wang C \emph{et al} 2009 Europhysics Lett. \textbf{86} 47002
\bibitem{Okada}Okada H \emph{et al} 2008 J. Phys. Soc. Jpn. \textbf{77} 113712
\bibitem{Rotter}Rotter M \emph{et al} 2008 Phys. Rev. B \textbf{78} 020503(R)
\bibitem{DaiJH}Dai J H, Si Q M, Zhu J X and Abrahams E 2009 Proc. Natl. Acad. Sci. \textbf{106}
4118
\bibitem{Izumi}Izumi F \emph{et al} 2000 Mater. Sci. Forum. \textbf{321-324} 198
\bibitem{Bao}Huang Q \emph{et al} 2008 Phys. Rev. Lett. \textbf{101} 257003
\bibitem{Si}Si Q and Abrahams E 2008 Phys. Rev. Lett. \textbf{101} 076401
\bibitem{Yildirim}Yildirim T 2008 Phys. Rev. Lett. \textbf{101} 057010
\bibitem{Dong}Dong J \emph{et al} 2008 Europhysics Lett. \textbf{83} 27006
\bibitem{PD122Co}Wang X F \emph{et al} 2009 New J. Phys. \textbf{11} 045003
\bibitem{P-SrBa122}Alireza P L \emph{et al} 2008 J. Phys.: Condensed Matter \textbf{21} 012208
\bibitem{Kimber}Kimber S A \emph{et al} 2009 Nat. Mater. \textbf{8}
471
\bibitem{Moriya}Moriya T and Ueda K 2003 Rep. Prog. Phys. \textbf{66} 1299

\end{thebibliography}
\end{document}